\documentclass[conference]{IEEEtran}
\usepackage[left=0.55in,right=0.55in,bottom=0.74in,top=0.74in]{geometry}
\renewcommand{\baselinestretch}{0.97} 
\usepackage{indentfirst}
\usepackage{cite}
\usepackage[tbtags]{amsmath}
\usepackage{amsfonts}
\usepackage{amssymb}
\usepackage{dsfont}
\usepackage{times}
\usepackage{subfigure}
\usepackage[english]{babel}
\usepackage{stfloats}
\usepackage{bbm}
\usepackage{dsfont}
\usepackage[dvips]{graphicx}
\usepackage{array}
\usepackage[normalem]{ulem}
\usepackage{epsf}
\usepackage{psfrag}
\usepackage{comment}
\usepackage{color}
\usepackage{xcolor}

\newcommand{\pr}{\mathcal{P}}

\newcounter{mytempeqncnt}

\begin{document}

\title{Modeling And Control Battery Aging in\\Energy Harvesting Systems}
\author{\large Roberto Valentini$^{* \dagger}$, Nga Dang$^{*}$, Marco Levorato$^{*}$, Eli Bozorgzadeh$^{*}$\\
\normalsize $*$ The Donald Bren School of Information and Computer Science, UC Irvine, CA, US\\
\normalsize $\dagger$ Dept. of Information Engineering, Computer Science and Mathematics, University of L'Aquila, IT \\
\normalsize e-mail: \{rvalent1,~levorato,~ngad,~ebozorgz\}@uci.edu\vspace{-0.6cm}}
\date{}

\maketitle

\pagestyle{empty}
\thispagestyle{empty}

\begin{abstract}
Energy storage is a fundamental component for the development of sustainable and environment-aware
technologies. One of the critical challenges that needs to be overcome is preserving the State of Health (SoH)
in energy harvesting systems, where bursty arrival of energy and load may severely degrade the battery. Tools from Markov process and Dynamic Programming theory are becoming 
an increasingly popular choice to control dynamics of these systems due to their ability to seamlessly incorporate heterogeneous components and support a wide range of applications.
Mapping aging rate measures to fit within the boundaries of these tools is non-trivial. In this paper, 
a framework for modeling and controlling the aging rate of batteries based on Markov
process theory is presented. Numerical results illustrate the tradeoff between battery degradation and task completion delay enabled by the proposed framework.
\end{abstract}

\vspace{-0.2cm}
\section{Introduction}
\label{sec:intro}

Energy storage systems represent a promising solution for smooth and robust integration of renewable energy sources to tomorrow's smart grid. In local micro-grid systems, the integration of energy storage  has been proposed not only as an effective way to buffer the high peaks of energy demand (load) to the grid, but also to smooth out the uncertainty and fluctuations which characterize renewable energy sources. Furthermore, energy storage solutions are also proposed for electrical vehicles to achieve higher energy efficiency \cite{dixon2010energy}, self-sustainable communication devices \cite{neely}, and cyber physical systems powered by renewable energy sources \cite{dang2015orchestrated}.

Among various technologies for energy storage, rechargeable batteries such as lithium-ion batteries are the prominent energy storage solution thanks to their relatively low cost and ability to hold charge. The main drawback of this technology is the limited battery life time and power density. Batteries cannot be charged and discharged an unlimited number of times due to aging effect. Their State of Health (SoH) not only depends on charge/discharge cycle counts (battery lifetime), but are also depends on charge/discharge rate (power density). As a consequence, the aging effect in batteries is even more dominant when deployed in renewable energy systems. Harnessing energy {\em opportunistically} from renewable sources causes a dynamic fluctuation in the charge level which may significantly degrade their SoH. In such systems, if the battery aging effect is not limited, the effective capacity and energy density may rapidly deteriorate. This paper proposes a novel modeling and optimization framework to capture and control the aging rate of batteries in these critical systems.

A considerable research effort is undergoing to develop models
for the degradation of the SoH of batteries over time. This is motivated by the possibility to design Energy Harvesting Systems (EHSs) relying on batteries with extended lifetime. In~\cite{mill}, the author proposes a degradation
model for lithium batteries, which is used in~\cite{date} and~\cite{date1} to optimize battery usage under aging
constraints. These works demonstrate that the SoH degradation rate can be quantified as a function of the battery usage in terms of the
average energy amount stored in the battery, named as average SoC and its standard deviation
over a time window. 

Based on these works, in this paper we present a novel framework for the modeling and control of the SoH of batteries
in EHSs. Different from prior work, the proposed framework models the temporal evolution of the system as
a Stochastic Finite State Machine (SFSM)~\cite{book_taylor}. This modeling rationale is becoming increasingly popular in a  
variety of Smart Grid related applications~\cite{o2010residential,6160649,levorato2012fast,jiang2011dynamic,tischer2011towards,
goonewardena2012charging,Kouts,levorato2014consumer}. For instance in~\cite{o2010residential}, a SFMS model capturing the dynamics of appliance activation and energy scheduling for residential demand response is presented. To minimize the weighted sum of the average financial cost of operations and appliance activation delay, the authors proposed a reinforcement learning approach. In~\cite{Kouts}, a Markov chain framework is proposed to find the optimal storage control policy for smarts power grids. In~\cite{neely}, the optimal utility scheduling for energy harvesting networks is analyzed. However, in these works, battery aging was not considered. The popularity of frameworks based on a SFSM/Markovian representation over other
options comes from the inherent simplicity of the model, which captures key dependencies in the temporal evolution
of the system, as well as interdependencies in the evolution (\emph{e.g.}, activation, de-activation) of individual components.
This representation also enables the use of a wide range of well studied analysis and optimization tools such
as dynamic programming~\cite{bertsekas1995dynamic} and hidden Markov models~\cite{elliott1995hidden}.

Herein, we develop a model and control framework that explicitly includes battery aging metrics in the optimization problem determining the behavior of the system. In~\cite{zorzi}, he authors propose a stochastic Markov chain optimization approach, where the age of the battery is explicitly included in the systems� state. However, the temporal scale of battery degradation is much larger than systems' time scale operations.
As a consequence, the approach in~\cite{zorzi} results in a coarse
approximation of battery aging. In this paper, we take a different approach by including metrics controlling the \emph{rate of aging} in the optimization problem. 

In order to make the scope of the proposed methodology as broad as possible, we consider a general EHS. The proposed formulation and metrics can be directly plugged into dynamic programming frameworks~\cite{bertsekas1995dynamic}.
The considered system allows to explore the tension between the minimization of the waiting time of energy tasks
generated, and stored, by the load module, and the need to avoid excessive SoC fluctuations. In particular, we show that SoH degradation can be reduced at the cost of a higher task completion delay. However, admitting this increased system delay may be necessary to avoid dramatic long-term system performance degradation in some cases.

The rest of this paper is organized as follows. In Section~\ref{sec:sysmod}, we introduce the SFSM system model. The battery aging modeling and definition of the associated cost functions are provided in Sections~\ref{sec:aging}. The optimization framework is described in Section~\ref{sec:opt}. In Section~\ref{sec:numres} numerical results are presented. Section~\ref{sec:conc} concludes the paper.
\section{System Model}
\label{sec:sysmod}
\begin{figure}[t!]
\centering
\includegraphics[scale=.3]{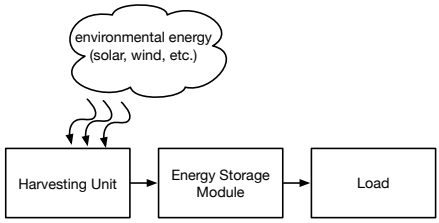}
\caption{Energy Harvesting System considered in this paper.}
\label{fig:fig0}
\end{figure}

We consider a general system composed of three main modules (see Fig.~\ref{fig:fig0}): the energy harvesting unit, the energy storage module, and the load module. The energy harvesting unit collects and converts energy from the environment to power the downstream modules. The energy storage device stores the energy acquired by the energy harvesting module, and interfaces with the load module, which models the arrival and queueing of energy requests from the application.

The system depicted in Fig.~\ref{fig:fig0} is instantiated as a SFSM with Markovian transitions. Slotted time is assumed,
where the index $\tau {\in} \mathbb{Z}^{+}$ indicates the time interval $[\tau \Delta T ,\tau \Delta T+\Delta T)$, where
$\Delta T$ is the slot duration. The overall FSM results from the composition of sub-FSMs associated with
the individual components (energy harvesting, battery and load). 

\begin{figure}[t!]
\centering
\includegraphics[scale=.3]{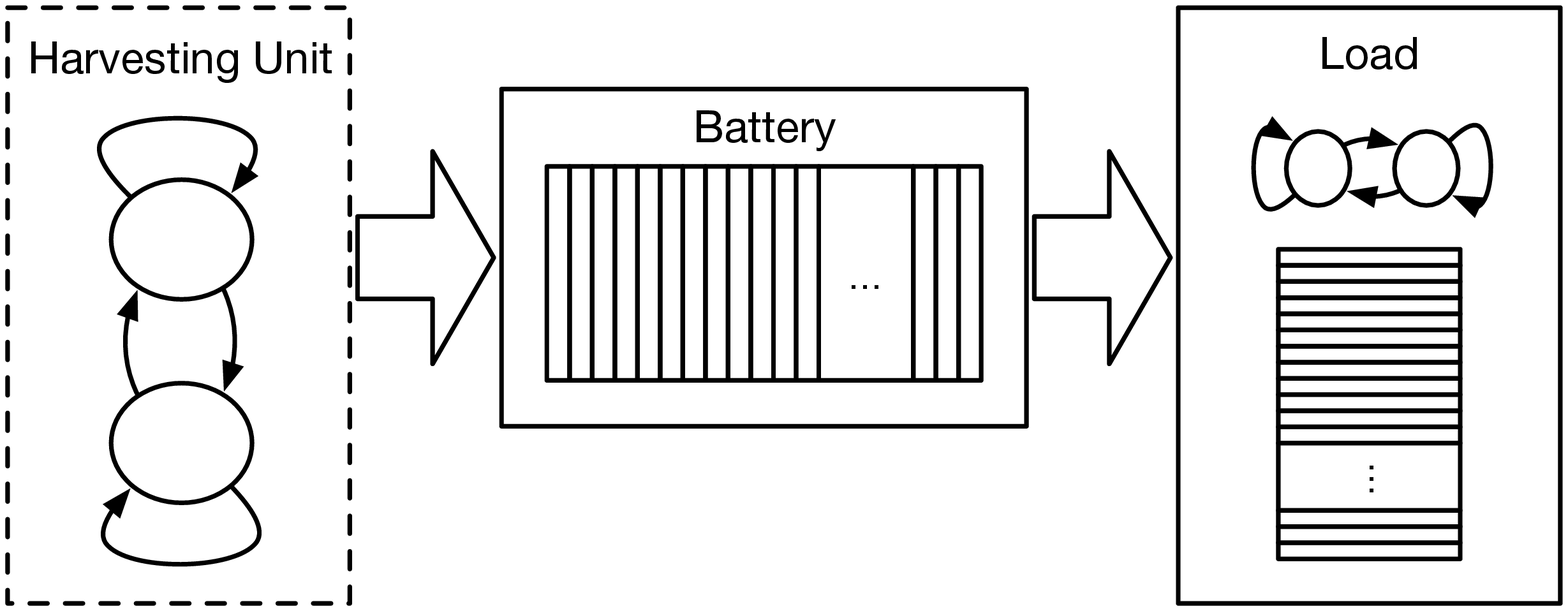}
\caption{System Model.}
\label{fig:fig1}
\end{figure}

Fig.~\ref{fig:fig1} illustrates the components of the system interpreted as FSMs.
The arrival of energy at the harvesting unit is modeled as the Markov process ${\bf H}{=}\{H_0,H_1,\ldots\}$, where 
$H_{\tau}{\in}\mathcal{H}$ is the harvesting state at time $\tau$, and $\mathcal{H}$ is the finite state space.
Define $E_{\tau}$ as the number of energy units entering the system at time $\tau$. Each state $h$ in $\mathcal{H}$
is associated with a distribution $f_{h}(e)$, where 
\begin{equation}
f_{h}(e){=}\pr(E_{\tau}{=}e|H_{\tau}{=}h).
\end{equation}
The temporal evolution of the harvesting process is governed by the transition probabilities
\begin{equation}
p_{\gamma_h}(h'{\mid}h){=}\pr(H_{\tau+1}{=}h'{\mid}H_{\tau}{=}h,\gamma_h),
\end{equation}
where $\gamma_h$ is a parameter set. Note that the sequence of energy unit arrivals $\{E_{0},E_1,\ldots\}$ is not a Markov process.

The battery is modeled as an ``energy buffer'' with nominal capacity $Q_{\rm max}$. The SoC is uniformly quantized with quantization step $\Delta Q$. Thus, the charge level at slot $\tau$, $Q_{\tau}$, takes values in the finite set $\mathcal{Q}{=}\{0,\ldots,Q_{\rm max}\}$. The temporal dynamics of the SoC are determined by the update equation
\begin{align}
\label{eq:q}
Q_{\tau+1}{=}\max\{Q_{\tau} {-} A_{\tau}, 0\} + E_{\tau},
\end{align}
where $A_{\tau}$ indicates the number of energy units used in slot $\tau$. As clarified later, $A_{\tau}{\in}\mathcal{A}{=}\{0\}\cup\{A_{\rm min},\ldots, A_{\rm max}\}$
is the ``action'' variable to be optimized. We assume that any combination of 
variables in the right hand side of Eq.~(\ref{eq:q}) generates an element in $\mathcal{Q}$. Alternatively, $\max\{Q_{\tau} {-} A_{\tau}, 0\} + E_{\tau}$
can be mapped to the closest element in $\mathcal{Q}$.

The load is modeled as a sequence of energy tasks to be completed. The energy tasks are stored in a finite buffer, and their arrival
is governed by a Markov process similar to that of energy arrival. Then, we define the Markov process ${\bf L}{=}\{L_0,L_1,
\ldots\}$, where $L_{\tau}{\in}\mathcal{L}$ is the state of the load in slot $\tau$ and $\mathcal{L}$ is the load state space. The temporal
evolution of the load state is determined by the transition probabilities 
\begin{equation}
p_{\gamma_l}(l'{\mid}l){=}\pr(L_{\tau+1}{=}l'{\mid}L_{\tau}{=}l,\gamma_l),
\end{equation}
where $\gamma_{l}$ is a parameter set. Each state $l{\in}\mathcal{L}$
is associated with the generation of an energy task whose completion necessitates a number of energy units $U$ determined by the
distribution $f_{l}(U)$, where
\begin{equation}
 f_{l}(u){=}\pr(U_{\tau}{=}u | L_{\tau}{=} l),
\end{equation}
and $U_{\tau}$ is the number of energy units requested at time $\tau$. The amount of energy units that has been requested, and not yet 
deployed, in slot $\tau$ is equal to $W_{\tau}{\in}\mathcal{W}{=}\{1,\ldots,W_{\rm max}\}$. The update rule of $W_{\tau}$, then, is
\begin{equation}
\label{eq:w}
W_{\tau+1}{=}\max(0,W_{\tau}{-}A_{\tau}) + U_{\tau}.
\end{equation}
Analogously to the update of the battery SoC, we build the system so that any combination of variables in the right hand side of Eq.~(\ref{eq:w}) generates an element in $\mathcal{W}$.
The above transition probabilities and update rules can be used to define the transition probabilities of the three modules as a function of the action distribution as described later in this paper.

We remark that the general FSM defined above can be adapted to capture the dynamics of systems with a higher complexity without the need for any major modification of the battery aging metrics and framework. For instance, the integration in the model of deadlines for the completion of energy tasks necessitates the inclusion of time counters in the state space representation of the load module.

\section{Battery Aging and Cost Functions}
\label{sec:aging}
Due to battery usage, the maximum charge level $Q_{\rm max}$ of the battery decays over time. Denote with
$Q_{\rm max}(t)$ the battery capacity at time $t$, then  $Q_{\rm max}(t) {\leq} Q_{\rm max}(t-1)$, and $Q_{\rm max}(0) {=} Q_{\rm max}$. We adopt the battery degradation model proposed in~\cite{mill} for continuous time systems, in which the rate of SoH degradation over
a time interval $[0,T]$, is a function of the average SoC, standard deviation of the SoC and the effective number of cycles, defined as
\begin{align}
SoC_{avg}&= \frac{1}{T}\int_{0}^{T} SoC(t) dt, \label{SoC_avg}\\
SoC_{dev}&= 2 \sqrt{\frac{3}{T}\int_{0}^{T} (SoC(t)-SoC_{avg})^{2} dt}, \label{SoC_dev}\\
N_{\rm cyc}&= \frac{1}{2Q_{\rm norm}}\int_{0}^{T} \mid I(t) \mid dt\label{N_cyc},
\end{align}
respectively, where $I(t)$ is the charging/discharging current of the battery. Based on the above metrics, the battery degradation
$D_{\eta, T}{=}f (SoC_{avg}, Soc_{dev}, N_{\rm cyc})$ in the target period, then, is computed as follows
\begin{align}
D_{\eta,T}&=D'~C~e^{D(SoC_{avg}-0.5)}, \label{eq:deg1}\\
D'&{=}A~N_{\rm cyc}~e^{(SoC_{dev}{-}1)B} + 0.2 \frac{T}{T_{life}}, \label{eq:deg2}
\end{align}
where $A$,$B$,$C$,$D$ and $T_{life}$ are constants associated with physical properties of the battery~\cite{mill}. However, different aging models can be easily incorporated in the optimization framework described in the next section. For instance, in~\cite{lam2013practical} the authors provide an empirical demonstration that the variable $N_{\rm cyc}$ has not a significant effect. In our framework, such modification only requires the removal of a constraint in the optimization problem.

Thus, battery aging exponentially increases with $SoC_{avg}$ and $SoC_{std}$, while linearly increases with $N_{\rm cyc}$.
Hence, in order to avoid high SoH degradation, the system needs to operate in low-charging levels and avoid SoC fluctuations with respect to the average SoC.
We remark that the time scale of battery aging is much larger compared to the time scale of the system's operations. Thus, rather than including the SoH in the state representation, our framework aims at minimizing the \emph{aging rate} over long time periods.

In order to apply dynamic programming techniques, the metrics defined in Eqs.~(\ref{SoC_avg}),~(\ref{SoC_dev}),
and~(\ref{N_cyc}) need to be transformed into time averages of additive cost functions mapped onto the state-action space of
the system. To obtain a representation with manageable complexity, this should be achieved by adding the smallest possible number of states to the state space. For the system described in Sec.~\ref{sec:sysmod}, the
discretized version of the metrics is 
\begin{align}
Q_{\mu} &= \frac{1}{T} \sum_{\tau=0}^{T} Q_{\tau}, \label{eq:c1}\\
Q^{2}_{\sigma} &= \frac{1}{T} \sum_{i=0}^{T} {(Q_{\tau} -Q_{\mu})}^{2},\label{eq:c2}\\
N_{\rm cyc} &= \frac{1}{2Q_{\rm nom}} \sum_{\tau=0}^{T} | A_{\tau} | + | E_{\tau}|,\label{eq:c3}
\end{align}
where $T$ is the number of considered slots.

While the metrics in Eqs.~(\ref{eq:c1}) and~(\ref{eq:c3}) find a direct transposition in the desired form, the metric in Eq.~(\ref{eq:c2}) requires a priori knowledge of the average SoC. Unfortunately, the average SoC is itself a function of the control, and $Q^{2}_{\sigma}$ does not find a direct representation
compatible with dynamic programming cost functions. To overcome this issue, we define a set of metrics whose minimization corresponds
to the minimization of $Q^{2}_{\sigma}$. Define, then, 
\begin{figure}[t!]
\centering
\includegraphics[scale=.38]{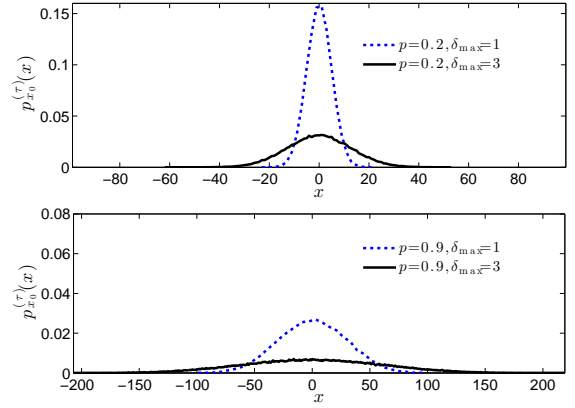}
\caption{Limiting distribution of the charge level $p^{(\tau)}_{x_0}(x)$ as a function of deviation from mean charge, for different values of the persistence parameter $p$ and the maximum step amplitude $\delta_{\rm max}$}
\label{fig:gauss}
\end{figure}
\begin{align}\label{eq:c4}
\Delta &= \frac{1}{T} \sum_{\tau=0}^{T} |\Delta_{\tau}|,\\
\label{eq:c5}
V &= \frac{1}{T-1} \sum_{\tau=1}^{T} \mathds{1}( {\rm sgn} (\Delta_{\tau}) {=} {\rm sgn}(\Delta_{\tau-1}) ),
\end{align}
where $\Delta_{\tau}{=}Q_{\tau}-Q_{\tau-1}$, $\mathds{1}(\cdot)$ the indicator function and ${\rm sgn(\cdot)}$ is the sign
function. Note that $\Delta $, represents the average amplitude of battery charge/discharge phases, while $V$ measures the average duration of a charge phase until battery starts discharging and conversely.

It can be proved that bounding $\Delta$ and $V$ is equivalent to bounding the SoC standard deviation. Due to
space limitations, we only provide a sketch of the proof. The SoC process can be modeled as a 
one-dimensional \emph{persistent random walk}~\cite{renshaw1981correlated}, for which the next step is taken in the same direction of the previous with probability $p$. Furthermore, consider a step amplitude $\delta$, uniformly distributed in $[1,\delta_{\rm max}]$.
The $\tau$-step distribution conditioned on the initial state $Q_0{=}x_0$, that is, $p^{(\tau)}_{x_0}(x){=}\pr(Q_{\tau}{=}x|Q_0{=}x_0)$,
converges to a Gaussian distribution when $\tau$ tends to infinity and the continuos limit is considered. The second moment of the limiting distribution, $\sigma^{2}{=}\bar{\delta}\tau p/(1-p)$, is a function of $p$ and the average step amplitude $\bar{\delta}$. Thus, by controlling the persistence $p$ and $\bar{\delta}$, which in our model corresponds to the control of $V$ and $\Delta$ respectively, the standard deviation of the process can be bounded. Fig.~\ref{fig:gauss} illustrates these interdependency between the cost metrics and the SoC variance.

\section{Optimization Framework}
\label{sec:opt}
We formulate an optimization problem aiming at minimizing the completion delay of the energy tasks with a bounded aging rate of the battery. Following the approach in~\cite{o2010residential}, we define the auxiliary variable
\begin{align}
Y_{\tau {+} 1}= \theta~ Y_{\tau} +(1{-}\theta)~W_{\tau},
\end{align}
with $ 0 {<} \theta {<} 1$. $Y_{\tau}$ exponentially smooths the task energy backlog and, thus, represents an indicator
of the average waiting time until all pending tasks are accomplished. When $\theta{=}0$, then $Y_{\tau+1}{=}W_{\tau}$ and $Y_{\tau}$ measures
the instantaneous load queue length. Note that $Y_{\tau}$ is a real valued variable, thus, we discretize the auxiliary variable to maintain model tractability, assuming that it takes values on a discrete finite set $\mathcal{Y}$.

The optimization problem can be formulated as a Constrained Markov Decision Process (CMDP)~\cite{ross1989randomized} aiming
at the minimization of the objective function
\begin{align}\label{eq:obj}
C = \lim_{T \rightarrow \infty} \frac{1}{T}E \left[ \sum_{\tau=0}^{T} \mathit{f}(Y_{\tau}) \right],
\end{align}
where $\mathit{f}(\cdot)$ is a convex function, under constraints on the following time averages
\begin{align}
Q_{\mu} &= \lim_{T\rightarrow\infty} \frac{1}{T}E\left[\sum_{\tau=0}^{T}Q_{\tau}\right], \label{eq:avcharge} \\
N_{\rm cyc} &=  \lim_{T\rightarrow\infty} \frac{1}{T}E\left[\sum_{\tau=0}^{T}| A_{\tau} | + | E_{\tau}|\right], \label{eq:avcyc} \\
\Delta &= \lim_{T\rightarrow\infty} \frac{1}{T}E\left[\sum_{\tau=0}^{T}| \Delta_{\tau} |\right], \label{eq:avdelta} \\ 
V &= \lim_{T\rightarrow\infty} \frac{1}{T}E\left[\sum_{\tau=0}^{T} g(\Lambda_{\tau})\right]. \label{eq:avpers} 
\end{align}

The definition of the cost function in Eq.~\eqref{eq:avpers} necessitates the inclusion in the model of the logical variable $\Lambda_{\tau}$, where $\Lambda_{\tau}{=}1$ if $\Delta_{\tau}{>}0$, corresponding to instantaneous (i.e., slot-by-slot) charging, and $\Lambda_{\tau}{=}0$ if $\Delta_{\tau}{>}0$, corresponding to instantaneous discharging.
By introducing a function $g(\cdot)$, such that $g(\Lambda_{\tau}){=}1$, if $\Lambda_{\tau}{=}{\Lambda_{\tau-1}}$, and $g(\Lambda_{\tau}){=}0$ otherwise, we measure the duration of charging and discharging phases.

Note that the objective function and constraints are defined as time averages over an infinite time horizon. This
formulation originates from the assumption that battery aging occurs at a time scale much larger than that
of the system's operations.

The overall system state $Z_{\tau}$ at time $\tau$ is defined to be the composition of all \emph{physical} sub-FSMs (i.e., harvesting, charge level, load), and the auxiliary variable $Y_{\tau}$ and the flag $\Lambda_{\tau}$, which are \emph{logical} sub-FSMs required to measure the average tasks' completion delay and battery degradation. In particular:
\begin{align}
Z_{\tau} = (H_{\tau},Q_{\tau},W_{\tau},Y_{\tau},L_{\tau},\Lambda_{\tau}),
\end{align}
with $Z_{\tau}{\in}\mathcal{Z}{=}\mathcal{H}{\times}\mathcal{Q}{\times}\mathcal{W}{\times}\mathcal{Y}{\times}\mathcal{L}{\times}\{0,1\}$. The transition probabilities are defined as $p(z' {\mid} z,a){=}\pr(Z_{\tau}{=}z' {\mid} Z_{\tau}{=}z,A_{\tau}{=}a)$, whit $z',z{\in}\mathcal{Z}$, $a{\in}\mathcal{A}$ and can be factorized as in Eq.~\eqref{eq:transprob}, given conditional independence between the various system components.\addtocounter{equation}{1}  

Without any loss of optimality, we constrain our search within the class of past-independent randomized policies $\mu{\colon}\mathcal{Z} {\times} \mathcal{A}{ \rightarrow}[0,1]$~\cite{ross1989randomized}, where $\mu(a{\mid}z){=}\pr(A_{\tau}{=}a{\mid}Z_{\tau}{=}z)$ is the probability of choosing action $a$ given that the system is in state $z$. Thus, the long-run objective in \eqref{eq:obj} can be redefined as a function of states and actions:
\begin{align}\label{eq:optim}
C_{\mu} = \sum_{z \in \mathcal{Z}}\sum_{a \in \mathcal{A}}\pi_{\mu}(z,a)c(z,a)
\end{align}
where $\pi_{\mu}(z,a){=}\pr_{\mu}(Z_{\tau}{=}z{,}A_{\tau}{=}a)$ is the stationary joint distribution of states and actions, $E_{\mu}[\cdot]$ is the expectation operator under policy $\mu$ and $c(z,a)$ is the value of the objective function incurring in state $z$ under action $a$.

Similar arguments hold for the time averages in Eqs.~\eqref{eq:avcharge},~\eqref{eq:avcyc},~\eqref{eq:avdelta}, and~\eqref{eq:avpers}, that can be mapped into functions of states and actions, $d^{k}{\colon}\mathcal{Z}{\times}\mathcal{A}{\rightarrow}\mathbb{R}$, with $k{=}\{1,2,3,4\}$. Each long-run average cost can be expressed as
\begin{align}
D_{\mu}^{k} =\sum_{z \in \mathcal{Z}}\sum_{a \in \mathcal{A}}\pi_{\mu}(z,a)d^{k}(z,a).
\end{align}
Due to space constraints, we do not include a detailed de- scription of the intuitive mapping between time averages and state-action functions.

The optimization problem can be formulated as
\begin{align}\label{eq:optproblem}
\begin{aligned}
& \underset{\mu}{\text{minimize}}
& & C_{\mu} \\
& \text{s.t.}
& & D_{\mu}^{k} \leq \hat{c}^{k}, \; k = 1, \ldots, 4.
\end{aligned}
\end{align}
where $\hat{c}^{k}$ is the upper bound for average cost of type $k$. This optimization problem is equivalent to the following linear program~\cite{ross1989randomized},
\begin{align}
& \underset{\mu}{\text{minimize}} 
& & \sum_{z \in \mathcal{Z}}\sum_{a \in \mathcal{A}}x_{za}c(z,a) \\
& \text{s.t.}
& & \sum_{a \in \mathcal{A}} x_{za} - \sum_{z \in \mathcal{Z}} \sum_{a \in \mathcal{A}} x_{za} p(z'{\mid}z,a) = 0 \\
& & &\sum_{z \in \mathcal{Z}}\sum_{a \in \mathcal{A}}x_{za}d^{k}(z,a) \leq \hat{c}^{k}, \; k = 1, \ldots, 4 \label{eq:agingcostr}\\
& & & \sum_{z \in \mathcal{Z}}\sum_{a \in \mathcal{A}}x_{za}=1 \\
& & & x_{za} \geq 0,
\end{align}
where the decision variables $x_{za}{=}\pi(z,a)$ are the steady state joint distributions of the states and actions. Then, the optimal stationary policy $\mu^{\star}$ can be computed as
\begin{align}
\mu^{\star}(a{\mid}z) = \frac{x_{za}}{\sum_{a\in \mathcal{A}}x_{za}}.
\end{align}

\section{Numerical Results}
\label{sec:numres}
In this section, we compare the performance of the aging-aware and age-unaware policies, and examine the tradeoff between the waiting time of backlogged energy tasks and SoH degradation. In particular, the resulting long-run average objective function and SoH degradation are analyzed for the constrained problem in Eq.~\eqref{eq:optproblem}, and the associated unconstrained version.
\begin{figure*}[b]
\normalsize
\setcounter{mytempeqncnt}{\value{equation}}
\hrulefill
\setcounter{equation}{23}
\begin{align}
\begin{aligned}
\label{eq:transprob}
p(z'{\mid}z,a)&= p_{\gamma_{h}}(h'{\mid}h)~p(q'{\mid}q,a)~p(w'{\mid}w,a)~p(y'{\mid}y,w)~p_{\gamma_{l}}(l'{\mid}l)~p(\lambda'{\mid}q',q)\\
&h,h'{\in}\mathcal{H},~q,q'{\in}\mathcal{Q},~w,w'{\in}\mathcal{W},~y,y'{\in}\mathcal{Y},~l,l'{\in}\mathcal{L},~\lambda'{\in}\{0,1\}.
\end{aligned}
\end{align}
\setcounter{equation}{\value{mytempeqncnt}}
\end{figure*}

We consider binary Markovian input processes $\mathbf{H}$ and $\mathbf{L}$, that is, $H_{\tau}{\in}\{0,1\}$ and $L_{\tau}{\in}\{0,1\}$. Furthermore, we assume that when $H_{\tau}{=}0$, no energy units arrive in the system, otherwise only one energy unit is generated with probability one. This corresponds to set $\pr(E_{\tau}{=}e|H_{\tau}{=}0){=}0$ and $\pr(E_{\tau}{=}1|H_{\tau}{=}1){=}1$. A similar model is used for the task arrival process. 

The transition probabilities for the input processes are defined by the parameter set $\gamma_{i}{=}(\phi_{i},b_{i})$, where $\phi_{i}$ and $b_{i}$ are the arrival rate and the average length of sequences of state ``1'' respectively. The index $i{=}h,l$, corresponds to the harvesting process ($h$) or the task arrival process ($l$). For the considered arrival model, the arrival rate corresponds to the steady state probability of state ``1''. Consequently, we have 
\begin{align}
p_{\gamma_{i} }(0 {\mid} 1) = \frac{1}{b_{i}}, ~~~ p_{\gamma_{i}} (1 {\mid} 0) = \frac{\phi_{i}}{b_{i}(1-\phi_{i})}.
\end{align}
In the numerical results, we set $\phi_{h}{=}0.9$ and $b_{h}{=}10$, meaning that the harvesting module frequently injects long bursts of energy into the system. The load arrival process has parameters $\phi_{l}{=}0.8$ and $b_{l}{=}12$.
\begin{figure}[t!]
\centering
\includegraphics[scale=.35]{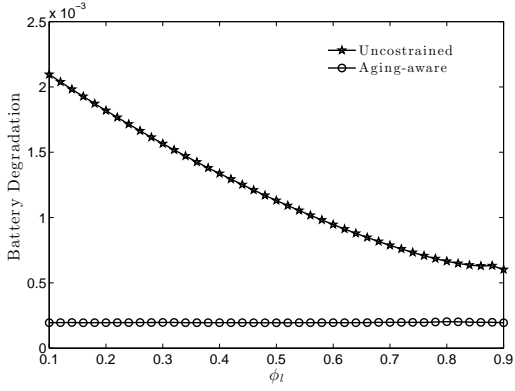}
\caption{Battery degradation as a function of tasks arrival rate $\phi_{l}$}
\label{fig:res1}
\end{figure}
\begin{figure}[t!]
\centering
\includegraphics[scale=.35]{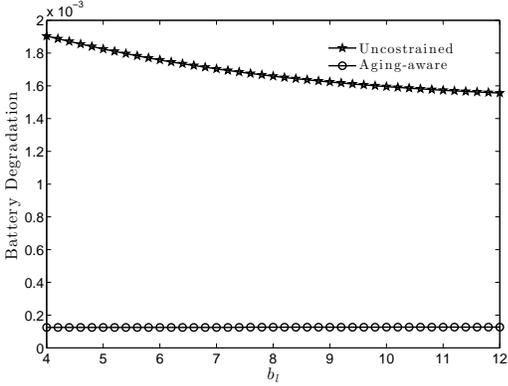}
\caption{Battery degradation as a function of task arrival process burstiness $b_{l}$}
\label{fig:res7}
\end{figure}
SoH degradation is computed based on the parameters of the battery cell for Plug-in Hybrid Electric Veichles (PHEVs) considered in~\cite{mill}, where the corresponding values for constants  $A$,$B$,$C$,$D$ and $T_{life}$ are provided.
For computational purposes we set $Q_{\rm max}{=}W_{\rm max}{=}8$ and we assume binary action space $\mathcal{A}{=}\{0,1\}$. Thus, the load can only remains ``idle'' or draws at most one energy unit from the battery that is sufficient for the completion of a task. 
Finally, in order to guarantee strict convexity of the objective function, we set $f(Y_{\tau}){=}Y_{\tau}^{2}$ and we choose $\theta{=}0.1$.

As shown in Fig.~\ref{fig:res1}, the aging-aware control policy guarantees low and almost constant degradation rate for different values of $\phi_{l}$ and $b_{l}{=}10$ with respect to the unconstrained policy. Note that when the aging constraints are removed, degradation rate decreases as the tasks arrival rate increases. This can be justified noting that, for fixed harvesting conditions, high values of $\phi_{l}$ correspond to frequent task arrivals, and consequently, frequent energy usage, leading to small values of $SoC_{avg}$, thus inducing a lower degradation rate.

Fig.~\ref{fig:res7} shows the aging rate as a function of the burstiness of the task arrival process $b_{l}$. Degradation is less sensitive to $b_{l}$ variations and slightly decreases as the load burstiness increases. Also in this case, the battery aging-aware approach guarantees very low and almost constant degradation rate.
\begin{figure}[t!]
\centering
\includegraphics[scale=.35]{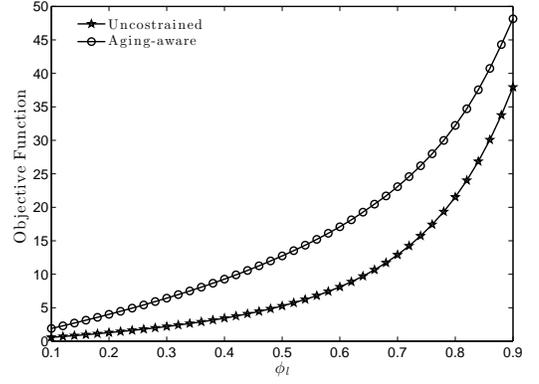}
\caption{Objective function over tasks arrival rate $\phi_{l}$}
\label{fig:res2}
\end{figure}
\begin{figure}[t!]
\centering
\includegraphics[scale=.35]{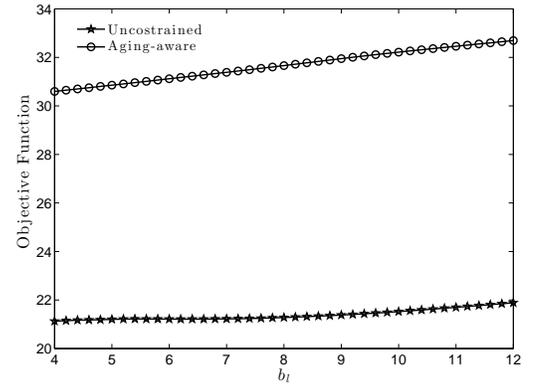}
\caption{Objective function over task arrival process burstiness $b_{l}$}
\label{fig:res3}
\end{figure}

Bounding the energy degradation by setting the constraints defined before generates a larger task-completion time. This is shown in Fig.~\ref{fig:res2} where, for every value of the tasks arrival rate, the aging-aware policy always results in higher delay. Note also that, as expected, the average waiting time increases as the load arrival rate increases.

For completeness, in Fig.~\ref{fig:res3} we also show the long-run average task completion delay as a function of the load burstiness $b_{l}$. We found that delay is less sensitive to burstiness of the load. However, the average delay induced by the constrained policy is still higher than that resulting from the unconstrained policy.

\begin{figure}[t!]
\centering
\includegraphics[scale=.35]{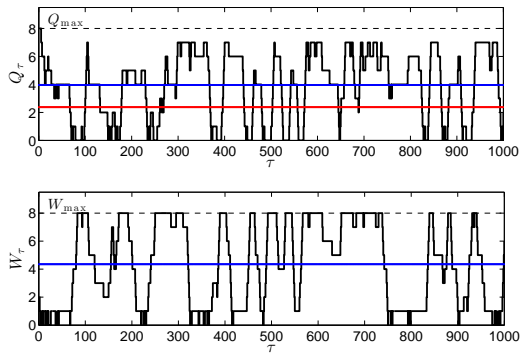}
\caption{Trace simulation under unconstrained policy. Heavy load, $\phi_{l}=0.8$, $b_{l}=12$. The blue line represents average and the red line represents standard deviation}
\label{fig:res5}
\end{figure}

\begin{figure}[t!]
\centering
\includegraphics[scale=.35]{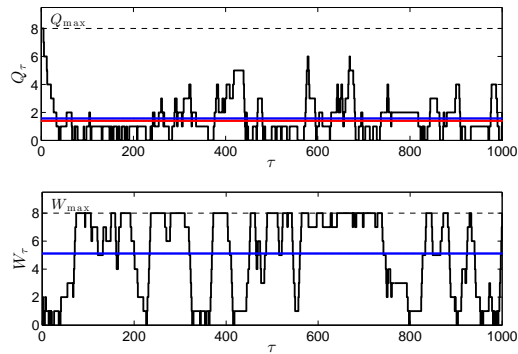}
\caption{Trace simulation under aging-aware policy. Heavy load, $\phi_{l}=0.8$, $b_{l}=12$. The blue and red lines represent average and standard deviation, respectively.}
\label{fig:res6}
\end{figure}

In order to provide some intuition on the structure of the control policies, we ran $1000$ samples Monte Carlo simulations both for the constrained and the unconstrained optimal policy. The resulting  charge level and task backlog traces are depicted in Fig.~\ref{fig:res5} (unconstrained policy) and Fig.~\ref{fig:res6} (aging-aware policy). The load buffer is assumed to operate under ``heavy load'' conditions, in particular we set $\phi_{l}{=}0.8$ and $b_{l}{=}12$.
As expected, the unconstrained policy results in a high number of charge/discharge cycles characterized by frequent and larger fluctuations around the mean value, whereas the aging-aware policy forces the battery to operate in a low  $SoC_{avg}$ regime characterized by smaller fluctuations around the mean, which is in turn kept bounded. Consequently the $SoC$ profile results smoother. This guarantees a low SoH degradation rate, thus prolonging battery lifetime and increasing system reliability. 
However, the aging-aware policy results in higher average backlogs with buffer saturation condition for $35.1\%$ of the time, against $28.2\%$  of the unconstrained policy.

\section{Conclusions}
\label{sec:conc}

In this paper, we presented a stochastic modeling and control framework for battery aging in EHS based on SFSM. Inspired by an existing battery degradation model, we introduced performance and aging measures that can be directly plugged into a CMDP in form of average cost functions. A linear program was proposed to solve the optimization problem, whose solution is the optimal randomized policy minimizing the average waiting time of energy tasks stored in a system queue with bounded degradation rate. Numerical results were provided to illustrate the tradeoff between system performance and battery degradation. 
\bibliographystyle{IEEEtran}
\bibliography{SMARTGRIDCOM_finalsubmission_ml.bbl}

\begin{thebibliography}{10}
\providecommand{\url}[1]{#1}
\csname url@rmstyle\endcsname
\providecommand{\newblock}{\relax}
\providecommand{\bibinfo}[2]{#2}
\providecommand\BIBentrySTDinterwordspacing{\spaceskip=0pt\relax}
\providecommand\BIBentryALTinterwordstretchfactor{4}
\providecommand\BIBentryALTinterwordspacing{\spaceskip=\fontdimen2\font plus
\BIBentryALTinterwordstretchfactor\fontdimen3\font minus
  \fontdimen4\font\relax}
\providecommand\BIBforeignlanguage[2]{{%
\expandafter\ifx\csname l@#1\endcsname\relax
\typeout{** WARNING: IEEEtran.bst: No hyphenation pattern has been}%
\typeout{** loaded for the language `#1'. Using the pattern for}%
\typeout{** the default language instead.}%
\else
\language=\csname l@#1\endcsname
\fi
#2}}

\bibitem{dixon2010energy}
J.~Dixon, ``Energy storage for electric vehicles,'' in \emph{IEEE International
  Conference on Industrial Technology (ICIT)}.\hskip 1em plus 0.5em minus
  0.4em\relax IEEE, 2010, pp. 20--26.

\bibitem{neely}
L.~Huang and M.~Neely, ``Utility optimal scheduling in energy-harvesting
  networks,'' \emph{IEEE/ACM Transactions on Networking}, vol.~21, no.~4, pp.
  1117--1130, Aug 2013.

\bibitem{dang2015orchestrated}
N.~Dang, H.~Tajik, N.~Dutt, N.~Venkatasubramanian, and E.~Bozorgzadeh,
  ``Orchestrated application quality and energy storage management in
  solar-powered embedded systems,'' in \emph{16th International Symposium on
  Quality Electronic Design (ISQED)}.\hskip 1em plus 0.5em minus 0.4em\relax
  IEEE, 2015, pp. 227--233.

\bibitem{mill}
A.~Millner, ``Modeling lithium ion battery degradation in electric vehicles,''
  in \emph{IEEE Conference on Innovative Technologies for an Efficient and
  Reliable Electricity Supply (CITRES)}, 2010, pp. 349--356.

\bibitem{date}
Q.~Xie, X.~Lin, Y.~Wang, M.~Pedram, D.~Shin, and N.~Chang, ``State of health
  aware charge management in hybrid electrical energy storage systems,'' in
  \emph{Design, Automation Test in Europe Conference Exhibition (DATE)}, 2012,
  pp. 1060--1065.

\bibitem{date1}
Y.~Wang, X.~Lin, Q.~Xie, N.~Chang, and M.~Pedram, ``Minimizing state-of-health
  degradation in hybrid electrical energy storage systems with arbitrary source
  and load profiles,'' in \emph{Design, Automation and Test in Europe
  Conference and Exhibition (DATE)}, 2014, pp. 1--4.

\bibitem{book_taylor}
H.~M. Taylor and S.~Karlin, \emph{An introduction to stochastic
  modeling}.\hskip 1em plus 0.5em minus 0.4em\relax John Wiley \& Sons, 1994.

\bibitem{o2010residential}
D.~O'Neill, M.~Levorato, A.~Goldsmith, and U.~Mitra, ``Residential demand
  response using reinforcement learning,'' in \emph{First IEEE International
  Conference on Smart Grid Communications (SmartGridComm)}, 2010, pp. 409--414.

\bibitem{6160649}
K.~Turitsyn, S.~Backhaus, M.~Ananyev, and M.~M.~Chertkov, ``Smart finite state
  devices: A modeling framework for demand response technologies,'' in
  \emph{50th IEEE Conference on Decision and Control and European Control
  Conference}, Dec 2011, pp. 7--14.

\bibitem{levorato2012fast}
M.~Levorato and U.~Mitra, ``Fast anomaly detection in smartgrids via sparse
  approximation theory,'' in \emph{IEEE 7th Sensor Array and Multichannel
  Signal Processing Workshop (SAM)}, 2012, pp. 5--8.

\bibitem{jiang2011dynamic}
B.~Jiang and Y.~Fei, ``Dynamic residential demand response and distributed
  generation management in smart microgrid with hierarchical agents,''
  \emph{Energy Procedia}, vol.~12, pp. 76--90, 2011.

\bibitem{tischer2011towards}
H.~Tischer and G.~G.~Verbic, ``Towards a smart home energy management system-a
  dynamic programming approach,'' in \emph{IEEE PES Innovative Smart Grid
  Technologies Asia (ISGT)}, 2011, pp. 1--7.

\bibitem{goonewardena2012charging}
M.~Goonewardena and L.~B. Le, ``Charging of electric vehicles utilizing random
  wind: A stochastic optimization approach,'' in \emph{2012 IEEE Globecom
  Workshops}.\hskip 1em plus 0.5em minus 0.4em\relax IEEE, 2012, pp.
  1520--1525.

\bibitem{Kouts}
I.~Koutsopoulos, V.~Hatzi, and L.~Tassiulas, ``Optimal energy storage control
  policies for the smart power grid,'' in \emph{IEEE International Conference
  on Smart Grid Communications (SmartGridComm),}, 2011, pp. 475--480.

\bibitem{levorato2014consumer}
M.~Levorato, N.~Ahmed, and Y.~Zhang, ``Consumer in-the-loop: Consumers as part
  of residential smart energy systems,'' in \emph{IEEE International Conference
  on Smart Grid Communications (SmartGridComm)}, 2014, pp. 758--763.

\bibitem{bertsekas1995dynamic}
D.~P. Bertsekas, \emph{Dynamic programming and optimal control}.\hskip 1em plus
  0.5em minus 0.4em\relax Athena Scientific Belmont, MA, 1995, vol.~1, no.~2.

\bibitem{elliott1995hidden}
R.~J. Elliott, L.~Aggoun, and J.~B. Moore, \emph{Hidden Markov Models}.\hskip
  1em plus 0.5em minus 0.4em\relax Springer, 1995.

\bibitem{zorzi}
N.~Michelusi, L.~Badia, R.~Carli, L.~Corradini, and M.~Zorzi, ``Energy
  management policies for harvesting-based wireless sensor devices with battery
  degradation,'' \emph{IEEE Transactions on Communications}, vol.~61, no.~12,
  pp. 4934--4947, December 2013.

\bibitem{lam2013practical}
L.~Lam and P.~Bauer, ``Practical capacity fading model for li-ion battery cells
  in electric vehicles,'' \emph{Power Electronics, IEEE Transactions on},
  vol.~28, no.~12, pp. 5910--5918, 2013.

\bibitem{renshaw1981correlated}
E.~Renshaw and R.~Henderson, ``The correlated random walk,'' \emph{Journal of
  Applied Probability}, pp. 403--414, 1981.

\bibitem{ross1989randomized}
K.~Ross, ``Randomized and past-dependent policies for markov decision processes
  with multiple constraints,'' \emph{Operations Research}, vol.~37, no.~3, pp.
  474--477, 1989.

\end{thebibliography}

\end{document}